\documentclass{article}

\usepackage{epsf}      
\usepackage{feynmp}    
\usepackage{amssymb}   
\usepackage{cite}

\skip\footins 39pt plus 4pt minus 2pt
\def\footnoterule{\kern-19pt\hrule width.5in\kern18.6pt}

\newcommand{\eq}[1]{equation~(\ref{#1})}
\newcommand{\Eq}[1]{Equation~(\ref{#1})}
\newcommand{\eqs}[2]{equations~(\ref{#1}) and~(\ref{#2})}

\newcommand{\fig}[1]{fig.~(\ref{#1})}
\newcommand{\Fig}[1]{Fig~(\ref{#1})}
\newcommand{\figs}[2]{figs~(\ref{#1}) and~(\ref{#2})}

\newcommand{\GLam}{G_{\Lambda}}
\newcommand{\GLamh}{G_{{\Lambda h}}}
\newcommand{\Gh}{G_{{h}}}

\newcommand{\SigmaLam}{\Sigma_{\Lambda}}
\newcommand{\SigmaLamh}{\Sigma_{{\Lambda h}}}

\newcommand{\ILam}{I_{\Lambda}}
\newcommand{\ILamh}{I_{{\Lambda h}}}

\newcommand{\be}{\begin{equation}}
\newcommand{\ee}{\end{equation}}

\newcommand{\bea}{\begin{eqnarray}}
\newcommand{\eea}{\end{eqnarray}}

\begin{document}

\renewcommand{\topfraction}{0.99}
\renewcommand{\bottomfraction}{0.99}
\twocolumn[\hsize\textwidth\columnwidth\hsize\csname 
@twocolumnfalse\endcsname

\begin{center}
{\Large \bf Fast Evaluation of Feynman Diagrams}  \\
\bigskip
{\large Richard Easther, Gerald Guralnik, and Stephen Hahn} \\
\smallskip
{\small Department of Physics,  Brown University, 
Providence, RI  02912, USA.}
\medskip

\begin{abstract}
We develop a new representation for the integrals associated with
Feynman diagrams. This leads directly to a novel method for the
numerical evaluation of these integrals, which avoids the use of Monte
Carlo techniques.  Our approach is based on based on the theory of
generalized sinc ($\sin(x)/x$) functions, from which we derive an
approximation to the propagator that is expressed as an infinite sum.
When the propagators in the Feynman integrals are replaced with the
approximate form all integrals over internal momenta and vertices are
converted into Gaussians, which can be evaluated analytically.
Performing the Gaussians yields a multi-dimensional infinite sum which
approximates the corresponding Feynman integral.  The difference
between the exact result and this approximation is set by an
adjustable parameter, and can be made arbitrarily small.  We discuss
the extraction of regularization independent quantities and
demonstrate, both in theory and practice, that these sums can be
evaluated quickly, even for third or fourth order diagrams.  Lastly,
we survey strategies for numerically evaluating the multi-dimensional
sums.  We illustrate the method with specific examples, including the
the second order sunset diagram from quartic scalar field theory, and
several higher-order diagrams. In this initial paper we focus upon
scalar field theories in Euclidean spacetime, but expect that this
approach can be generalized to fields with spin.

\end{abstract}

\end{center}

BROWN-HET-1158

\bigskip
\mbox{}
]


\section{Introduction} 

The expansion of relativistic quantum field theory developed over 50
years ago by Feynman and Schwinger is still the basis for all
perturbative calculations.  However, analytical evaluations of graphs
have become increasingly sophisticated, and digital computers make it
possible to calculate graphs numerically.  Since the accuracy of
experiments in fundamental particle physics is continually improving,
obtaining theoretical predictions whose precision matches that of the
data requires the evaluation of ever larger numbers of more complex
graphs.  Moreover, non-Abelian and effective field theories also
increase the variety of diagrams that must be considered.  Thus the
evaluation of Feynman diagrams remains an active area of research.

Many Feynman diagrams can be partially, or completely, evaluated with
purely analytical methods.  When a diagram cannot be reduced to a
closed form, the remaining integrals must be tackled numerically.
Since these integrals may be multi-dimensional, Monte Carlo
integration is usually the only practical numerical algorithm
\cite{Lepage1978a,PressBK1}.  While conceptually simple, Monte Carlo
methods converge slowly and accurate calculations of non-trivial
graphs can be extremely time consuming.  However, this combination of
analytical and numerical techniques is extremely powerful and, for
instance, makes it possible to evaluate all the eighth order graphs in
QED which contribute to the electron magnetic moment
\cite{Kinoshita1990a}. In addition, computer algebra methods can be
used to tackle the analytical categorization and simplification of
diagrams, as summarized in the recent review
\cite{HarlanderET1998a}.

In this paper, our immediate goal is to present a new approach to the
numerical evaluation of Feynman diagrams.  Its principal advantages
are that it applies to arbitrary topologies, involves a minimal amount
of analytic overhead, and has the potential to be much faster than
Monte Carlo methods, especially when a result with more than one or
two significant figures is required.  The numerical method we develop
is based on a novel representation of the Feynman integrals
themselves, which allows us to approximate an arbitrary Feynman
integral with a multi-dimensional infinite sum.

The basis of our approach is an approximation to the spacetime
propagator, derived using the theory of generalized Sinc functions and
expressed as an infinite sum.  The approximation is governed by a
small parameter, $h$, and converges rapidly to the exact result as
$h\rightarrow 0$. For our purposes, the crucial feature of the
approximate propagator is that its spatial (or momentum) dependence
appears in terms like $\exp{(-(x-y)^2)}$.  When the propagators inside
a Feynman integral are replaced with this approximate form, integrals
over vertex locations or internal momenta required by the Feynman
rules {\em all\/} reduce to Gaussian integrals, which can {\em all\/}
be performed explicitly for {\em any\/} diagram. Once the Gaussian
integrals have been performed, the result is an $N$-dimensional
infinite sum, where $N$ is the number of propagators in the diagram.
We refer to this infinite sum as the Sinc function representation of
the Feynman integral.

An immediate advantage of the Sinc function representation is that
sums are much easier to compute numerically than integrals, since sums
are intrinsically discrete.  More importantly, the general term in the
sum decays exponentially, which greatly facilitates its numerical
evaluation.  We show that the effort needed to evaluate the $N$
dimensional Sinc function representation of a Feynman integral
increases linearly with the desired number of significant figures,
although for a complicated topology the constant of proportionality
can be large.  This is in contrast to Monte Carlo methods where each
successive significant figure is generally more costly to obtain than
the last, even when the convergence is improved through the use of
adaptive sampling.

This paper introduces the Sinc function representation, and
demonstrates its use in several explicit examples.  We also present an
approach to regularizing divergent diagrams and discuss the extraction
of renormalization independent quantities in the context of the Sinc
function representation.  We focus our attention on scalar field
theories in Euclidean spacetime, and will address the extension of
Sinc function methods to more complicated theories in future work.

Finally, while the Sinc function representation applies to
perturbative quantum field theory, the approximate propagator is also
a key ingredient of a new approach to ``exact'' numerical quantum
field theory, the `source Galerkin' method
\cite{GarciaET1994a,GarciaET1996a,LawsonET1996a,%
Hahn1998a,HahnET1999a}.
The source Galerkin method also eschews the use of Monte Carlo
calculations. In addition to improving its computational efficiency,
by avoiding Monte Carlo techniques the source Galerkin method
sidesteps many of the problems faced by other numerical approaches to
non-perturbative quantum field theory, particularly in models with
fermions.  While the physical bases of the Sinc function
representation and the source Galerkin method are very different,
there is some intriguing evidence that insights gained while
constructing the Sinc function representation may also allow us to
improve the performance of the source Galerkin method.

The structure of this paper is as follows: In Section 2 we define the
generalized sinc functions, and review their relevant properties.  The
Sinc function representation of the propagator is described in Section
3.  In Section 4 we show this form of the propagator leads to a new
representation for Feynman integrals, and derive the ``Sinc function
Feynman rules'' for scalar field theory.  We then apply these rules to
the two loop ``sunset'' contribution to the $\lambda\phi^4$ two-point
function, and compare our results to a conventional analytic
calculation.  In Section 5 we discuss techniques for efficiently
evaluating the multi-dimensional sums obtained from the Sinc function
Feynman rules. We then evaluate the sums derived for representative
third and fourth order diagrams, and compare these calculations to
direct integrations with VEGAS.  Finally, in Section 6, we summarize
our results and identify questions which we will pursue in future
work.

\section{Generalized Sinc Functions}

We begin by describing a generalized Sinc function, which is defined by
\be
S_k(h,x) = \frac{\sin{\left[\pi (x-kh)/h\right]}}{\pi(x-kh)/h}.
\ee
This is an obvious extension of the usual sinc function,
$\mbox{sinc}(x) = \sin(x)/x$.  Stenger \cite{StengerBK1} gives a
thorough discussion of these functions, while a more introductory
account is provided by Higgins
\cite{Higgins1985a}. We follow Stenger in using the capitalized
``Sinc'' to distinguish the generalized version, although our notation
for $S_k(h,x)$ differs from his.\footnote{Our $S_k(h,x)$ is equivalent to
Stenger's $S(k,h)(x)$.}

The Sinc function has the integral representation,
\be
S_k(h,x) = \frac{h}{2 \pi} \int_{-\pi/h}^{\pi/h}{e^{\pm i (x-kh)t}\,dt},
\label{intrep}
\ee
which is trivial to prove. We recognize \eq{intrep} as the Fourier
transform of a finite wavetrain, which is non-zero only in the
interval $[-\pi/h,\pi/h]$.  Using the integral representation, we can
establish the orthonormality of the Sinc functions,
\begin{eqnarray}
\int_{-\infty}^{\infty}{dx\,S_k(h,x)} &=& h, \label{norm} \\
\int_{-\infty}^{\infty}{dx\,S_k(h,x) S_l(h,x)} &=& h \delta_{kl}.
\label{orthog}
\end{eqnarray}

Any function $f(z)$, which is analytic on a rectangular strip of the
complex plane, that is centered on the real axis with width $2d$, is
approximated by
\be
f(z) \approx \sum_{k=-\infty}^{\infty} f(kh) S_k(h,z).
\label{sincexpansion}
\ee
The approximation improves as $h$ decreases, and this relationship is
quantified below. \Eq{sincexpansion} is referred to as the ``Sinc
expansion'' or ``Sinc approximation'' of $f(z)$. The Sinc expansion is
exact for Paley-Weiner, or band-limited, functions, which corresponds
to $f(z)$ be representable in the following way:
\be
f(z) = \int_{-\pi/h}^{\pi/h}{g(u) e^{izu}\, du}, \qquad 
        g \in L^2(-\pi/h,\pi/h).
\ee
This states that $f$ is a function whose Fourier transform, $g$, has
compact support on the interval $[-\pi/h,\pi/h]$ or, alternatively,
that $f$ has no frequency outside the ``band'', $[-\pi/h,\pi/h]$.
Sinc functions are used in a wide variety of areas, and Stenger gives
a number of examples from applied mathematics and classical
physics. They are also at the basis of Shannon's Sampling Theorem and
are frequently encountered in communications theory.

For our purposes the most important property of the Sinc function is
that if $f$ is approximated by its Sinc expansion, \eq{sincexpansion},
we can quickly derive a related approximation for the definite integral,
\begin{eqnarray}
\int_{-\infty}^{\infty} {f(z)\, dz } &\approx& 
   \sum_{k=-\infty}^{\infty} \int_{-\infty}^{\infty}{dz\, f(kh) S_k(h,z)} 
\nonumber \\
 &=& h\sum_{k=-\infty}^{\infty} f(kh)
\label{sincintegral}
\end{eqnarray}
where the last equality follows from the normalization, \eq{norm}. In
the following section we use this relationship to approximate the
propagator as an infinite sum.

{}From Theorem~3.1.2 of Stenger, the error in the Sinc approximation
to a function $f(z)$ which is analytic in an infinite strip of the
complex plane, centered upon the real axis with width $2 d$, is
\begin{eqnarray}
\delta f(h,z) &=& f(z) - \sum_{k=-\infty}^{\infty} f(kh) S_k(h,z) \\
 &=& \frac{\sin{\frac{\pi z}{ h}}}{2 \pi i} 
   \int_{-\infty}^{\infty}  \left\{
  \frac{f(t- i d)}{(t-z-id)\sin\left[\frac{\pi}{h}(t- id)\right]}  \right.
\nonumber \\
 & & \qquad + \left.  
  \frac{f(t+ i d)}{(t-z+id)\sin\left[\frac{\pi}{h}(t+ id)\right]}
 \right\}\, dt  \label{errorf}.
\end{eqnarray}
The discrepancy between $\int_{-\infty}^{\infty}f$ and its Sinc
approximation is
\be
\Delta f(h) = \int_{-\infty}^{\infty}{dz\, f(z)} - 
  h \sum_{k=-\infty}^{\infty} f(kh). 
\ee
Clearly, $\Delta f(h)$ is the result of integrating $\delta f(h,z)$
between $\pm \infty$, and Theorem~(3.2.1) of Stenger allows us to deduce
that
\be
|\Delta f(h)| \le C \frac{e^{-\pi d/h}}{2 \sinh{\pi d/h}}
\label{Deltaf}
\ee
where $C$ is a real number which is independent of $h$. When  $h< d$
we see that $|\Delta f(h)|$ drops exponentially as $h$ decreases
linearly.

\section{The Scalar Field Propagator}

\subsection{The Spacetime Propagator}

Feynman diagrams are conventionally computed in momentum space, since
the propagators are algebraically simpler, and the momentum
conservation constraint at each vertex reduces the number of integrals
that would otherwise need to be performed. However, we will find it
more convenient to perform most of our work in coordinate space.

The spacetime propagator for a scalar field theory is the Fourier
transform of the momentum space propagator,
\be
G(p) = \frac{1}{p^2 + m^2} \label{Gp}
\ee
namely\footnote{This statement also implicitly defines the
normalization we have adopted for the Fourier transform and its
inverse. We also assume $m^2 >0 $.}
\be
G(x-y) = \int{\frac{d_E^4 p}{(2 \pi)^4} \frac{e^{i p (x-y)}}{p^2+m^2}},
\ee
where $x$, $y$ and $p$ are Lorentz four-vectors,\footnote{We will
suppress the Lorentz indices on four-vectors unless they are directly
relevant to the calculation at hand.} and we are working in Euclidean
space. Redefining $x-y$ as $x$ and adding a cut-off in anticipation of
divergent diagrams, we can write
\be
\GLam(x) = \int{\frac{d_E^4 p}{(2 \pi)^4}\, \frac{e^{i p x}}{p^2+m^2}
  e^{-p^2/\Lambda^2 }}.
\ee
Exponentiating the denominator gives
\be
\GLam(x) = \int{\frac{d_E^4 p}{(2 \pi)^4}\, e^{i p x - p^2 /\Lambda^2} 
        \int_{0}^{\infty} {ds e^{-s(p^2 +m^2) }}}.
\ee
A rationale for this regulated form was given by Hahn
\cite{Hahn1998a}; in particular, formulating the cut-off in this way
preserves the Gaussian form of the integrals (as would a
dimensionally-regulated formulation). Reversing the order of
integration, completing the square, performing the Gaussian integrals,
and finally shifting from $s$ to $s/m^2$ yields
\be 
\GLam(x) = \frac{m^2}{(4\pi)^2 }\int_{0}^{\infty} {ds\,
 \frac{1}{(s + \frac{m^2}{\Lambda^2})^2} 
  \exp{\left[-s - \frac{m^2 x^2}{4(s + \frac{m^2}{\Lambda^2})} \right]}}. 
\label{GLam1}   
\ee
Making a further change of variable to, $s = e^z$, we obtain
\bea
\GLam(x) &=& \frac{m^2}{(4\pi)^2 }\int_{-\infty}^{\infty} dz\, 
 \frac{e^z}{(e^z + \frac{m^2}{\Lambda^2})^2}  \nonumber \\
 & &  \qquad \times 
\exp{\left[-e^z - \frac{m^2 x^2}{4(e^z + \frac{m^2}{\Lambda^2})} 
  \right]}. \label{GLam2}
\eea
The crucial step is to approximate $\GLam(x)$ as an integral over a
Sinc expansion  \cite{Hahn1998a} of the form given by \eq{sincintegral}, 
\begin{eqnarray}
\GLamh(x) &\approx& \frac{m^2 }{(4\pi)^2 }\int_{-\infty}^{+\infty}
        dz\, \sum_{k=-\infty}^{\infty}{ 
 \frac{e^{k h}}{(e^{k h} + \frac{m^2}{\Lambda^2})^2} } \nonumber \\
 &\mbox{}& \qquad \times 
 \exp{\left[-e^{k h} - \frac{m^2 x^2}{4(e^{k h} + \frac{m^2}{\Lambda^2})}
  \right]} S_k(h,z) \nonumber \\
 &=& \frac{m^2 h}{(4\pi)^2 }\sum_{k=-\infty}^{\infty} 
 \frac{e^{k h}}{(e^{k h} + \frac{m^2}{\Lambda^2})^2} \nonumber \\
&& \qquad  \times \exp{\left[-e^{k h} - \frac{m^2 x^2}{4(e^{k h} +
\frac{m^2}{\Lambda^2})} \right]}. \label{GLamh2}
\end{eqnarray}
When $\Lambda \rightarrow \infty$ the coordinate space propagator,
$G(x)$, can be expressed as a modified Bessel function of the second
kind,
\be
G(x) = \frac{m^2}{4\pi^2} \frac{K_{1}(m|x|)}{m|x|}. \label{G1}
\ee
In this limit, the Sinc approximation to $\GLam$ also simplifies,
\be
\Gh(x) = \frac{m^2 h}{(4\pi)^2 }\sum_{k=-\infty}^{\infty} {
   \exp{\left[-k h - e^{k h} - \frac{m^2 x^2}{4e^{k h}} \right]}}. 
\label{Gh1}
\ee
Finally, it will be convenient to introduce a more concise
representation for the various factors that appear in the general term
of sum,
\be
\GLamh(a-b) =  \frac{m^2 h}{(4\pi)^2 }\sum_{k=-\infty}^{\infty}{ 
  p(k) \exp{\left[- \frac{m^2 (a-b)^2}{4 c(k)} \right]}}
\label{GLamh3}
\ee
where 
\begin{eqnarray}
c(k) &=& e^{kh} + \frac{m^2}{\Lambda^2}, \label{ckdef} \\
p(k) &=& \frac{e^{kh}\exp{(-e^{kh})}}{(e^{kh} + \frac{m^2}{\Lambda^2})^2}
         = \frac{\exp{(kh-e^{kh})}}{c(k)^2}. \label{pkdef}
\end{eqnarray}

We now have four different versions of the propagator, which are
related as follows. Firstly, the exact propagator, $G(x)$, is given by
\eq{G1}, while \eq{GLam1} defines the cut-off propagator, $\GLam(x)$.
These expressions are approximated by $\Gh(x)$ and $\GLamh(x)$
respectively, which are defined by \eqs{Gh1}{GLamh2}. As we shall see,
the approximations rapidly approach the exact values as $h$ becomes
small.

Both the cut-off propagator, $\GLam(a-b)$, and its approximation,
$\GLamh(a-b)$, depend only on the combination $(a-b)^2$, and thus
neither the introduction of $\Lambda$ nor the Sinc function
approximation has violated Poincar\'{e} invariance.  Moreover, by
Fourier transforming $\GLamh(x)$ we could obtain a Sinc function
approximation for the momentum-space propagator, \eq{Gp}, in the
presence of the cut-off.  In Section 4, we develop the Sinc function
Feynman rules in coordinate space, but the momentum space rules could
be obtained using an almost identical argument.

\begin{figure}[tbp]
\begin{center}
\begin{tabular}{c}
\epsfxsize=8cm 
\epsfbox{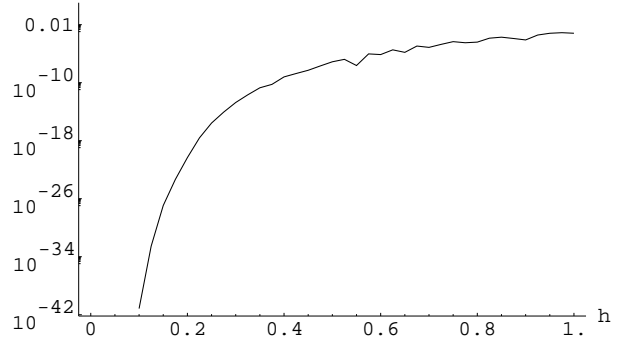}
\end{tabular}
\end{center}
\caption[fig1]{The relative error of the approximation $\Gh$ is shown, by
plotting $|\Delta \Gh / G|$ as a function of $h$ for $m=x=1$ and
$\Lambda\rightarrow\infty$.
\label{hdepends}}
\end{figure}

\begin{figure}[tbp]
\begin{center}
\begin{tabular}{c}
\epsfxsize=8cm 
\epsfbox{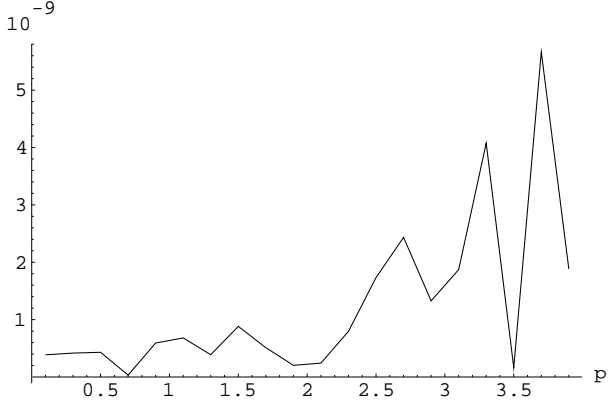}
\end{tabular}
\end{center}
\caption[fig1]{The relative error of the approximation $\Gh$ is shown, by
plotting $|\Delta \Gh / G|$ as a function of $x$ for $m=1$, $h=0.4$
and $\Lambda\rightarrow \infty$.
\label{xdepends}}
\end{figure}

\subsection{Accuracy of the Approximation}

Our next task is to assess the accuracy of the approximation involved in
writing $\Gh(x)$ and $\GLamh(x)$.  We define
\be
\Delta \GLamh(x) = \GLam(x) - \GLamh(x).
\ee
The Sinc function approximation to the propagator is derived from
\eq{sincintegral}, so the $h$ dependence of the error is governed by
\eq{Deltaf}. The form of $f(z)$ which appears in the left hand side of
\eq{Deltaf} is our integrand,
\be
f(z) = \frac{e^z}{(e^z + \frac{m^2}{\Lambda^2})^2} \exp{\left[-e^z -
\frac{m^2 x^2}{4(e^z +\frac{m^2}{\Lambda^2})} \right]}
\ee
where $x$, $m$ and $\Lambda$ are simply parameters. Since $f(z)$ is
infinite when $e^z + m^2 /\Lambda^2$ =0, or when $z= 2\log(m/\Lambda)
\pm i \pi$, it follows that $d$, the width of the strip of the complex plane 
which appears in \eq{errorf}, can take any value in the open interval
$(0,\pi)$. Using the maximal possible value of $d$, we derive the
following bound on the accuracy of the approximation,
\be
|\Delta \GLamh| \sim e^{-2\pi/h}, \qquad h\rightarrow 0.
\ee

Thus as $h$ approaches zero, $\Delta f(h,z)$ decreases
exponentially. Looking at the general form of the Sinc expansion, we
see that the magnitude of the general term in the sum decreases (at
least) exponentially with increasing $k$. Since any numerical
evaluation of these sums must be cut off at finite $|k|$, it follows
that while $\GLamh$ will approach $\GLam$ exponentially quickly as $h$
goes to zero, the number of terms that need be considered when
computing the sum grows (at worst) linearly with $1/h$. This
impressive convergence facilitates the numerical evaluation of the
approximate propagators and, consequently, the Sinc function
representations of the Feynman integrals.

Having discussed the abstract convergence properties of the Sinc
approximation to the scalar field propagator, we now present some
specific numerical results.  \Fig{hdepends} shows $\Delta\Gh(1)/G(1)$
as a function of $h$. These results were obtained using arbitrary
precision arithmetic. When $h=0.1$, the two expressions differ by less
than 1 part in $10^{-40}$, and if $h$ is even slightly less than
unity the accuracy of the approximation is phenomenally impressive. As
a specific example, choosing $h=0.5$ gives an accuracy of 1 part in
$10^8$, while $h = 0.25$ yields an agreement better than 1 part in
$10^{16}$. On most computers, double precision real numbers contain 16
decimal digits, so if $h\lesssim 0.25$, $\Gh(x)$ and $G(x)$ will be
numerically indistinguishable.  In \fig{xdepends} we show that for
fixed $h$, the accuracy of the approximation is essentially
independent of $x$.

In \figs{hdepends}{xdepends} we have taken $\Lambda\rightarrow\infty$
for convenience, since this allows us to use the exact expression for
the propagator, \eq{G1}. However, using a finite value of $\Lambda$
makes no significant difference to these conclusions.

\begin{figure}[tbp]
\begin{center}
\begin{tabular}{c}
\epsfxsize=8cm 
\epsfbox{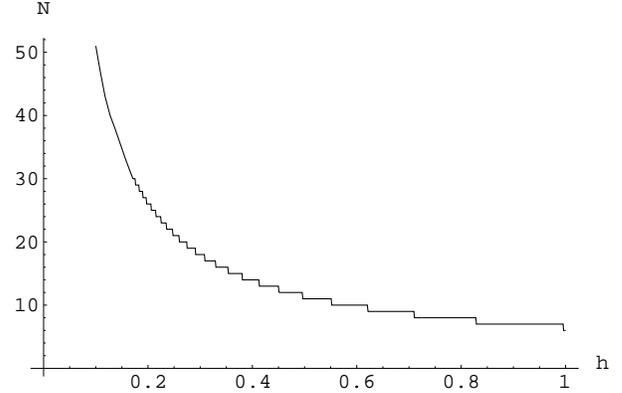}
\end{tabular}
\end{center}
\caption[fig2]{The number of significant terms in the series for
$\Gh(1)$, with $m=1$ is displayed. When the double infinite series for
$\Gh(x)$ is truncated at $k=-N$ and $k=N$, the result is accurate to
within approximately 1 part in $10^{14}$.  
between 
\label{converge}}
\end{figure}

Our next concern is to determine how many terms in the series
representation of $\Gh$ contribute significantly to the final result
when it is evaluated numerically. \Fig{converge} shows the number of
terms in the series for $\Gh$ which contribute more than 1 part in
$10^{14}$ of the final result.  If $h=0.25$, we need to include terms
for which $|k|<21$. Note that this measures the convergence to $\Gh$
rather than $G$, but for $h \lesssim 0.25$ the difference is less than
the truncation error in the numerical calculation of $\Gh(x)$.  Also,
if we do not need a highly accurate calculation, we can increase the
value of $h$ and thus reduce the number of terms that contribute to
the sum.

\section{Evaluating Feynman Diagrams}

\subsection{The Sinc Function Representation}

Having developed the Sinc function approximation to the propagator, we
now turn to the evaluation of Feynman diagrams. In coordinate space
the Feynman rules specify that we construct the product of the
propagators corresponding to the $N$ internal lines of the diagram,
and then integrate over each of the $M$ internal vertices.  If the
external lines are attached at the points $x_1,x_2,\cdots,x_A$ and the
internal vertices are labeled by $y_1,y_2,\cdots,y_M$ we are therefore
integrating over all spacetime for each of the $y_j$.  Up to overall
prefactors consisting of powers of the coupling constants and the
topological weight of the diagram, these integrals take the form
\bea
\ILam &=& \int d^4 y_1 
  \cdots d^4 y_M \GLam(x_1-y_1) \cdots   \nonumber \\
&& \quad \GLam(x_A-y_j) \cdots \GLam(y_k-y_M). \label{ILam1}
\eea
In general, the $\ILam$ cannot be reduced to a closed, analytic form.
We now use the Sinc expanded propagator, $\GLamh$, to derive the Sinc
function representation, $\ILamh$, of a Feynman integral, $\ILam$.

Firstly, $\GLam(x) = \GLamh(x) + \Delta \GLamh(x)$. For given
$\Lambda$ and $h$, there exists a number $c$, such that $\Delta \GLamh
(x) \lesssim c \GLam(x)$, and $c \ll 1$ if $h$ is sufficiently small.
We therefore obtain
\bea 
\ILam &=& \int d^4 y_1
\cdots d^4 y_M \left[ \right. (\GLamh(x_1-y_1) + \nonumber \\
&& \qquad \Delta \GLamh(x_1-y_1)) \cdots
\nonumber \\
&& \quad \left.  (\GLamh(y_k-y_M)+ \Delta \GLamh(y_k-y_M))
\right] \\
&\approx& \int d^4 y_1
\cdots d^4 y_M \GLamh(x_1-y_1) \cdots \GLamh(y_k-y_M) \nonumber \\
&& \qquad {} + M \int d^4 y_1
\cdots d^4 y_M \left[ \Delta\GLamh(x_1-y_1) \cdots \right. \nonumber \\
&& \qquad \left. \GLamh(y_k-y_M) \right], \eea
where we have assumed that $h$ is small enough to make $\Delta
\GLamh(x) \ll \GLamh(x)$, so we can drop terms that are second order
in the $\Delta \GLamh(x)$. Consequently, we can use the definition of
$c$ to write
\bea
\ILam &\approx& \int d^4 y_1\cdots d^4 y_M \GLamh(x_1-y_1) \cdots
\GLamh(y_k-y_M)
\nonumber \\ 
&& + c M \int d^4 y_1 \cdots d^4 y_M \GLamh(x_1-y_1) \cdots
\\ && \qquad \GLamh(y_k-y_M) \nonumber \\
&\approx& \ILamh \label{ILamh1}
\eea
where
\be
\ILamh = \int d^4 y_1 \cdots d^4 y_M \, \GLamh(x_1-y_1) \cdots \GLamh(y_k-y_M).
\label{ILamh2}
\ee
We see that the Sinc function representation, $\ILamh$, differs from
$\ILam$ by ${\cal O}(c)$. Since $c\rightarrow0$ as $h\rightarrow 0$,
$\ILamh$ is an arbitrarily accurate approximation to $\ILam$.

The spatial dependence of $\GLamh(y_1-y_2)$ occurs only in terms like
$\exp{(-(y_1-y_2)^2)}$. Thus the integrals over the internal vertices
in \eq{ILamh2} are all reduced to Gaussians, which can be performed
analytically, no matter how complicated the diagram.  Thus the Sinc
function representation is an $N$ dimensional infinite sum, where $N$
is the number of internal lines in the diagram. In general, this sum
has the form
\be
\ILamh= h^N \sum_{k} f(k_1,\cdots,k_N,x_1,\cdots,x_M,h), 
\ee
where the $\sum_{k}{}$ stands for the $N$ individual summations
between $\pm\infty$, corresponding to each of the $k_i$.  Referring
back to \eq{GLamh3}, we see that the $k_i$ only appear in $f$ inside
the $c(k_i)$ and $p(k_i)$ of the original propagators.  From the form
of $\GLamh$ and the properties of Gaussian integrals, it follows that
that the spatial dependence of $f$ is restricted to terms with the
form $\exp{(-(x_a-x_b)^2)}$. Consequently, like $\GLamh(x)$ the
$\ILamh(x)$ are also manifestly covariant.

For reference, a Gaussian integral in four (Euclidean) spacetime
dimensions has the general form
\be
\int{d^4 x \, \exp{\left( -a + 2b_\mu x^\mu - d x_\mu x^\mu \right)}}
= \frac{\pi^2 e^{-a}}{d^2} \exp{\left(\frac{b_\mu b^\mu}{d} \right)},
\label{gdef}
\ee
and we have actually already used this result to derive $\GLam(x)$.
Furthermore, the Fourier transform of a Gaussian,
\be
\int{d^4 x \, e^{-i p x} e^{-d x^2} } = \frac{\pi^2}{d^2}
 \exp{\left( -\frac{p^2}{4d}\right)},
\ee
is simply a special case of \eq{gdef}. Using this result, it is easy
to transform $\ILamh(x)$, and obtain the corresponding momentum space
representation.

Before we discuss the convergence properties and evaluation of these
sums, we note that the spatial integrations in $\ILam$ can also be
performed if we work with integral representation of the cut-off
propagator, \eq{GLam1}.  Consider \eq{ILam1},
\bea
\ILam &=& \left(\frac{m}{4\pi}\right)^{2N} \int_0^\infty{s_1\cdots s_N 
}\prod_{i=1}^N{\frac{e^{-s_i}}{(s_i + \frac{m^2}{\Lambda^2})^2} }
\times \nonumber \\ && \quad \int{d^y_1\cdots d^y_M \exp{\left(
-\frac{m^2 (x_1 -y_1)^2}{ 4(s_1 + \frac{m^2}{\Lambda^2})}\right)}}
\cdots \nonumber \\ && \qquad \exp{\left( -\frac{m^2 (y_k -y_M)^2}{
4(s_N + \frac{m^2}{\Lambda^2})}\right)} .  \label{ILam2}
\eea
Combining the Gaussians, we obtain an $N$-dimensional integral over
the parameters, $s_i$. It is this integral which is directly
approximated by the Sinc function representation, $\ILamh$.

Writing an $N$-dimensional integral as an $N$ dimensional sum is not
{\em necessarily\/} an improvement, since the sums still have to be
numerically evaluated.  Moreover, while $\ILamh$ is an $N$-dimensional
sum, other approaches express to the Feynman integrals with less than
$N$ dimensions.  For instance, if we had integrated over the exact
form of the scalar propagator in $4-\epsilon$ dimensions the result
would still be finite, but we only need to perform $M$ integrals over
the internal vertices.  Obviously, $M$ is always less than $N$, and
the Euclidean integrals can typically each be expressed as a single
integral over a radial variable, even though they are not Gaussians.
Likewise, in momentum space the dimension of the integral is
determined by the number of independent momenta, which must be also be
smaller than the number of propagators.  However, it will turn out
that the Sinc approximations to Feynman integrals inherit the rapid
convergence properties possessed by $\GLamh$, and even complicated
diagrams are computationally tractable.

At this point it is useful to consider a na\"{\i}ve estimate of the
computational cost of evaluating $\ILam$ using the infinite sum
derived from $\ILamh$.  If the diagram contains $N$ propagators, and
we truncate the individual sums at $|k_i| \sim n$, we would expect to
evaluate $\sim n^N$ terms. A complicated diagram can have $N\sim10$,
and in this case modest values of $n$ will still make $n^N$
exorbitantly high, even when we are evaluating millions of terms a
second.  However, in the next sections we show that this simple
calculation is actually far too pessimistic. Firstly, the number of
significant terms grows more slowly with $N$ than the volume of an
$N$-dimensional hypercube, and a sensible numerical algorithm will
focus on these terms.  More radically, we also consider approaches
which reduce the overall ``power'' of the problem to a number less
than $N$.

Even at this point, though, we can probe the relationship between the
accuracy of the approximation to $\ILam$ and the number of terms that
must be evaluated in $\ILamh$. Recall that $\Delta \GLamh$ decreases
exponentially with $h^{-1}$, while the number of terms significant
terms in the sum increases linearly.  Since the number $c$ that
appears in the derivation of \eq{ILamh1} is effectively the maximum
value of $\Delta \GLamh(x)$ for $0\le x^2<\infty$, $\ILamh$ inherits
the convergence properties of $\GLamh(x)$.  Consequently, to double
the number of significant digits of $\ILam$ given by $\ILamh$, we must
halve the value of $h$. Roughly speaking, this increases the number of
terms which contribute to $\ILamh$ by a factor of $\sim 2^N$.  This
$2^N$ may seem a little daunting, but the important point is that the
number of significant terms still grows {\em linearly\/} with the
desired accuracy.

It is this relationship between $h$ and the number of significant
terms in $\ILamh$ that holds out the promise of a dramatic improvement
over Monte Carlo methods. For comparison, while adaptive Monte Carlo
routines such as VEGAS focus on the most important parts of the
overall volume, obtaining high accuracy is inherently difficult
because of the statistical nature of the method. At worst, the
accuracy of the Monte Carlo algorithm scales as $1/\sqrt{n}$ where $n$
is the number of points to be evaluated. Since each decimal place in
the result increases the accuracy by a factor of 10, the CPU time
needed for each successive significant figure can be as much as 100
times greater than that required for its predecessor.  Thus the time
required by a Monte Carlo integration can rise exponentially with the
desired accuracy, in contrast with the linear increase that applies to
the Sinc function method.

\subsection{Renormalization and Sinc \\Function Methods}

The next issue we discuss is regularization and renormalization when
using the Sinc function representation.  Firstly, we wish to remove a
possible source of confusion by clarifying the two different limits
implied by $\ILamh$, namely $\Lambda\rightarrow\infty$ and
$h\rightarrow0$.  In the limit $h\rightarrow0$, the error implicit in
the Sinc function representation drops to zero. However, when we use
the Sinc function representation as the basis of a numerical
calculation, we do not (and cannot) evaluate this limit exactly.
Rather, we choose a finite value of $h$ that is small enough to ensure
that the error induced by the Sinc expansion is less than the accuracy
we require from our computation. Beyond this, the value of $h$ has no
physical significance.

In contrast, when $\Lambda\rightarrow\infty$ most non-trivial
integrals diverge, and the corresponding infinities must be removed.
Also, recall that while Feynman integrals can be regularized (rendered
finite) in many different ways, renormalized quantities cannot contain
any residual dependence on the regularization scheme.

In this initial paper we focus on diagrams which contribute to the
propagator.  The possible divergences of a diagram with no
sub-divergences scale as $p^0$ and $p^2$, so for a given $\Sigma(p)$
which contributes to the propagator we can form a renormalized
quantity,
\be
\tilde{\Sigma}(p) = \lim_{\Lambda \rightarrow \infty}{
 \left[\Sigma(p) - \Sigma(0) - p^2 \left. \frac{d
\Sigma(p)}{d(p^2)} \right|_{p=0}\right]}, \label{regprop}
\ee
where the two terms that have been dropped are simply the first two
contributions to the Taylor expansion of the regularized diagram,
$\Sigma(p^2)$. We can see no {\em a priori\/} reason not to use
dimensional regularization, but in practice we have found the
covariant, $\Lambda$-dependent cut-off to be convenient.  However, any
cut-off scheme which did not ensure that the integrals over the
internal vertices are always Gaussians would undermine the Sinc
function representation, since there would be no guarantee that the
spatial integrals in an arbitrary diagram could all be performed.

A given $\Sigma(p)$ which contributes to the propagator has a Sinc
function representation with the generic form
\be 
\Sigma(p) \sim \sum_{k}A \exp(-p^2 B) \label{format} 
\ee
where $A$ and $B$ are functions of the $k_i$, $\Lambda$, $h$, and $m$,
but independent of $p$. In principle, we could evaluate the
corresponding $\tilde{\Sigma}(p)$ by computing the terms on the right
hand side of \eq{regprop} at a large (but still finite) value of
$\Lambda$, and then performing the subtractions.  Unfortunately, this
na\"{\i}ve approach is foolish since the individual terms in
\eq{regprop} are dominated by the divergences and combining them would
lead to a drastic loss of numerical precision.

We resolve this problem by writing the right hand side of \eq{regprop}
as a single multi-dimensional sum,
\be
\tilde{\Sigma}(p) \sim  \sum_{k}\left[ \lim_{\Lambda\rightarrow \infty} 
  A(\exp(-p^2 B)  -1 + p^2 B) \right]. \label{regprop1}
\ee
Mathematically this is not objectionable, since by combining the
individual terms while keeping $\Lambda$ finite we never manipulate
formally divergent sums. We take the limit $\Lambda\rightarrow\infty$
in \eq{regprop1} analytically, {\em before \/} we compute the
multi-dimensional sum.  Not only does this avoid the loss of precision
inherent in the na\"{\i}ve evaluation of \eq{regprop}, it yields an
analytical, renormalized, cut-off independent expression for
$\tilde{\Sigma}(p)$.

The numerical task we face is to evaluate the sum that results from
taking the limit in \eq{regprop1}. In general, this is no more
difficult than evaluating the cut-off dependent expression,
\eq{format}. The one caveat is that if $p^2 B \ll 1$ the individual
terms in the summand of \eq{regprop1} have very similar magnitudes,
and combining them directly would lead to a loss of numerical
precision.  However, when $p^2 B \ll 1$ the combination we need can be
computed accurately by replacing the exponential with the first few
terms of its Taylor series, which converges very quickly.

We give a specific example which illustrates this approach to
renormalization when we compute the sunset diagram from $\lambda
\phi^4$. Moreover, we believe that this argument can be generalized to
diagrams with sub-divergences, but we will reserve a detailed
discussion of this problem for a future publication.

\subsection{Sinc Function Feynman Rules}

We are now in a position to write down the analog of the conventional
Feynman rules that specify how to obtain the Sinc function
representation for an arbitrary diagram in scalar field theory:
\begin{enumerate}
\item Write down the integral to be evaluated, using the usual
coordinate space Feynman rules and  propagators.
\item Replace the propagators with $\GLamh(x)$, the Sinc expansion
of the cut-off propagator. 
\item The spatial integrals are now reduced to Gaussians, which
are performed analytically.
\item Once the Gaussian integrals have been performed,
Fourier transform the result into momentum space (if desired).
\item Extract the regularization independent quantities (if desired).
\item Evaluate the resulting multi-dimensional sum numerically.
\end{enumerate}
Before proceeding, we make two observations.  Firstly, the Gaussian
integrals can easily be performed using computer algebra packages, and
the result expressed in the syntax of compiled languages such as
Fortran or C. This suggests that the derivation and evaluation of the
Sinc function representation for a given group of diagrams will be
easy to automate.  Secondly, while the general term in the sum will
grow more complicated as the number of internal vertices is increased,
the integrals that must be performed are always Gaussians.
Consequently, extracting the Sinc function representation is not
intrinsically more difficult for higher order diagrams than it is for
simple diagrams.

\begin{figure}
\begin{center}
\begin{fmffile}{sunsetfile}
\begin{fmfgraph*}(150,50)
  \fmfpen{thick}        
  \fmfleft{i1} %
  \fmf{fermion,label=$p$}{i1,v1}
  \fmf{fermion,label=$p$}{v2,o1}
  \fmf{vanilla}{v1,v2}
  \fmf{vanilla,left,tension=.1}{v1,v2,v1}
  \fmfright{o1} 
  \fmfdotn{v}{2}
\end{fmfgraph*}
\end{fmffile}
\end{center}
\caption[]{The  sunset diagram contribution to the $\lambda \phi^4$
propagator, with momentum, $p$. \label{sunsetdiag}}
\end{figure}
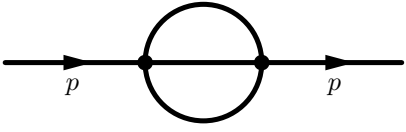

\begin{figure}[tbp]
\begin{center}
\begin{tabular}{c}
\epsfxsize=8cm 
\epsfbox{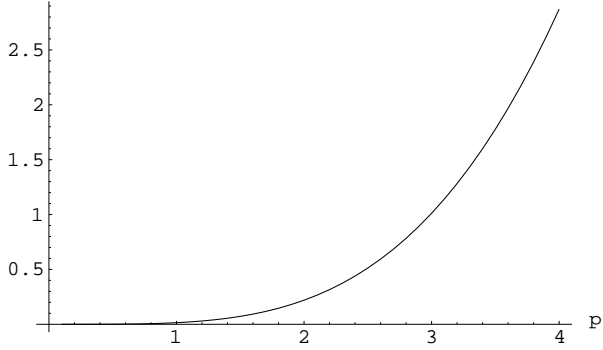}
\end{tabular}
\end{center}
\caption[]{The contribution of the sunset diagram to
$\tilde{\sigma}$ is plotted. The data shown here was obtained from the
Sinc approximation with $h=0.4$, but is indistinguishable from the exact
result. The initial factor of $m^2 /(4\pi)^4$ in $\tilde{\Sigma}(p)$ has
been omitted from these results.
\label{sunset1}}
\end{figure}

\subsubsection{Case Study I: The Sunset Diagram}

To give a specific example of our method, we evaluate the sunset
diagram which contributes to the propagator of $\lambda \phi^4$ scalar
field theory, and which is depicted in \fig{sunsetdiag}.  This diagram
is a useful test subject, since its overlapping divergences ensure
that it is has non-trivial structure, but it is still simple to enough
to be treated analytically.  Consequently, we can use it to to test
the Sinc function method to an arbitrary level of precision.

In coordinate space the diagram has a particularly simple form, since
there are no internal vertices to integrate across. Consequently, we
have
\be
\SigmaLamh(x-y) = \GLamh(x-y)^3
\ee
which is just the product of three propagators. We use the same
convention here as we do with $\GLamh$ and $\ILamh$, where the cut-off
dependent Sinc approximation is denoted by the subscripts $\Lambda$
and $h$.  We have suppressed the prefactor consisting of $\lambda^2$
and the topological weight, since it plays no part in the
analysis. Working through the prescription given above,
\bea
\SigmaLamh(x-y) &=& 
\frac{m^6 h^3}{(4 \pi)^6} \sum_{k}
p(k_1) p(k_2) p(k_3) \times \nonumber \\
&& \quad \exp\left[ -\frac{m^2 (x-y)^2}{4} \right. \nonumber \\
&& \qquad \left.\left(
\frac{1}{c(k_1)}  + \frac{1}{c(k_2)}  + \frac{1}{c(k_3)} \right)
\right],
\eea
After carrying out the Fourier transform, we obtain
\bea
\SigmaLamh(p) &=& 
\frac{m^2 h^3}{(4 \pi)^4} \sum_{k}
\frac{p(k_1) p(k_2) p(k_3)}{\left(
\frac{1}{c(k_1)}  + \frac{1}{c(k_2)}  + \frac{1}{c(k_3)} \right)^2}
  \nonumber \times \\
&& \quad \exp\left[ -\frac{p^2}{m^2}  \frac{1}{
\frac{1}{c(k_1)}  + \frac{1}{c(k_2)}  + \frac{1}{c(k_3)} }
\right].
\eea
We have now turned the integral represented by the sunset diagram into
an infinite sum over all possible $\{k_1,k_2,k_3\}$. Starting from
\eqs{GLam1}{ILam2}, we can derive the triple integral to which
$\SigmaLamh$ is the approximation,
\bea
\SigmaLam(p) &=& \frac{m^2}{(4\pi)^4} \int_0^\infty ds_1 ds_2 ds_3 
\prod_{i=1}^{3} \left( \frac{1}{\tilde{s}_i^2} e^{-s_i}\right) \times
\nonumber \\
&& \quad \frac{1}{\left( \tilde{s}_1^{-1} +
\tilde{s}_2^{-1}+\tilde{s}_3^{-1} \right)^2} \times \nonumber \\
&& \qquad
\exp{\left(-\frac{p^2}{m^2}  \frac{1}{ \tilde{s}_1^{-1} +
\tilde{s}_2^{-1}+\tilde{s}_3^{-1} }\right)},
\eea
where $\tilde{s}_i$ stands for $s_i + m^2 /\Lambda^2$.  The parallel
between $\SigmaLamh$ and $\SigmaLam$ is clear.

\begin{figure}[tbp]
\begin{center}
\begin{tabular}{c}
\epsfxsize=8cm 
\epsfbox{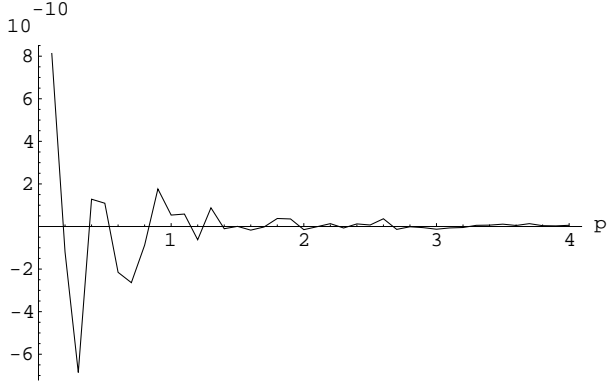}
\end{tabular}
\end{center}
\caption[]{The relative error in the Sinc function calculation of
the finite part of the sunset diagram, $\tilde{\Sigma}$, is shown,
compared to obtained from the analytical result, \eq{sunsetdim}, with
$h=0.4$.
\label{sunset2}}
\end{figure}

\begin{figure}[tbp]
\begin{center}
\begin{tabular}{c}
\epsfxsize=8cm 
\epsfbox{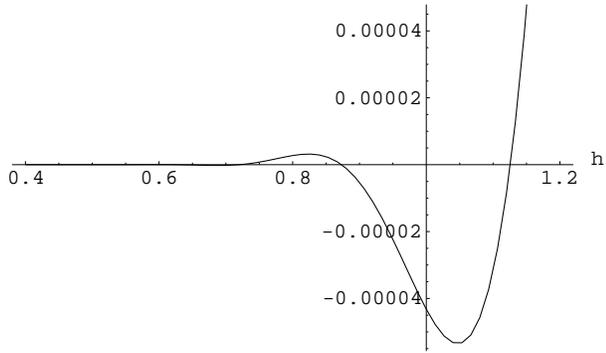}
\end{tabular}
\end{center}
\caption[]{The relative error in the Sinc function representation of
the sunset diagrams is plotted as a function of $h$, with a fixed $p$
of 1.4. Even when $h \sim 1$, the Sinc function method returns several
significant figures of the exact result, while the nested sum can be
evaluated in a fraction of a second.
\label{sunset3}}
\end{figure}

A little analysis establishes that both $\SigmaLam$ and $\SigmaLamh$
are finite, but that they diverge as $\Lambda \rightarrow \infty$,
which is precisely what we expect, since this diagram contains an
(overlapping) divergence. Analyzing the general term in the sum, we
see that the divergence is ``worst'' when we sum along the line
$k_1=k_2=k_3$, $k_1<0$. Applying the renormalization prescription
given by \eq{regprop1} we form
\bea
\tilde{\Sigma}_h(p) &=& 
\frac{m^2 h^3}{(4 \pi)^4} \sum_{k}
\frac{p(k_1) p(k_2) p(k_3)}{\left(
\frac{1}{c(k_1)}  + \frac{1}{c(k_2)}  + \frac{1}{c(k_3)} \right)^2} \times
  \nonumber \\
&& \quad \left( \exp\left[ -\frac{p^2}{m^2}  \frac{1}{
\frac{1}{c(k_1)}  + \frac{1}{c(k_2)}  + \frac{1}{c(k_3)} }
\right] \right. \nonumber \\
&& \qquad \left. -1 + \frac{p^2}{m^2}  \frac{1}{
\frac{1}{c(k_1)}  + \frac{1}{c(k_2)}  + \frac{1}{c(k_3)} }\right) .
\label{sunseth}
\eea
The result of evaluating \eq{sunseth} is plotted in \fig{sunset1}.

We now compare this result to a standard analytic computation. Rather
than tackle the evaluation of $\SigmaLam$ directly, we use the results
of Ramond \cite{RamondBK1}, who works in $4-2\epsilon$ dimensions to
compute that
\be
\Sigma(p) = \frac{-1}{1-2\epsilon} (\mu^2) ^{2 \epsilon} \left(
3m^2 K(p) + p^\mu K_\mu (p)\right) \label{sunsetdim}
\ee
where $\mu$ is the renormalization scale and the $K(p)$ and
$p^\mu K_\mu(p)$ can be expanded as functions of $\epsilon$ to give
\bea
K(p) &=&
\frac{\Gamma(2\epsilon)}{(4\pi)^{4-2\epsilon}}\frac{1}{\epsilon}
\left[ \rule{0pt}{18pt}
1 + \epsilon\left(1 - 2\log{m^2} \right)  + \right. \nonumber \\
 & &  \epsilon^2 \left( 3- \frac{\pi^2}{6} + 2\log^2(m^2) -
 4 \log(m^4) \right. \nonumber \\ 
&&  + 2 \int_0^1 dx \int_0^1 dy (1-y) \log(y) \frac{d \log(f)}{dy}
  \nonumber \\
&& \left. \left.
 - 2   \int_0^1 dx \int_0^1 dy \log(f) \log(y) \right)
\rule{0pt}{18pt} 
\right]
\label{Kp}
\eea
where
\be
f= p^2 y (1-y) + m^2 \left(1-y - \frac{y}{x(1-x)}\right).
\ee
In the two unevaluated integrals, a single integration can be
performed without too much difficulty, and the result expressed in
terms of dilogarithms.
In addition,
\bea
p^\mu K_\mu (p) &=&
\frac{\Gamma(2\epsilon)}{(4\pi)^{4-2\epsilon}}\frac{1}{\epsilon}
\left[ \frac{1}{2} - \epsilon\left(\frac{1}{4} + \log{(m^2)}
\nonumber \right. \right.\\
&& \left. \left. + 2 \int_0^1 dy (1-y) \beta
 \log{\frac{\beta +1}{\beta  -1}} \right)\right] \label{Kmup}
\eea
where
\be
\beta = \frac{m^2 y }{ p^2 (1-y) y + m^2(1-y)}.
\ee

Normally, infinities are eliminated from dimensionally regularized
diagrams by dropping terms proportional to $1/\epsilon^2$ and
$1/\epsilon$.  However, if we drop the first two terms in the Taylor
expansion with respect to $p^2$, and then take $\epsilon\rightarrow 0$
we remove the divergent terms, and also ensure that the result is (in
theory) equivalent to \eq{sunseth}, in the limit $h\rightarrow 0$.
The agreement between the exact result and the Sinc function
approximation is shown in \figs{sunset2}{sunset3}. As predicted, we
can obtain an arbitrarily good agreement with the exact result by
lowering $h$, but we also find that discrepancy between the
approximation and the exact result remains small even when $h$ is of
order unity.  Larger values of $h$ reduce the number of terms that
contribute significantly to the sum, and if we only want a few
significant figures we can set $h\sim1$, which makes the diagrams very
easy to evaluate numerically.

\section{Numerical Implementation}

\subsection{A Simple Approach}

So far, we have focused on the theoretical development of the Sinc
function representation.  In this section we turn our attention to the
practical issues that arise when numerically evaluating the sums, and
discuss the CPU time needed for typical calculations.

Na\"{\i}vely, computing a sum is a much simpler process than
evaluating an integral since a sum is inherently discrete, whereas an
integral is a continuous object which must be discretized before being
numerically evaluated.  The one proviso is that we must truncate the
sum carefully, to ensure that terms which do not contribute
significantly to the final result are not needlessly calculated, and
that terms which do not contribute are not accidentally ignored.

\begin{figure}[tbp]
\begin{center}
\begin{tabular}{c}
\epsfxsize=7.5cm 
\epsfbox{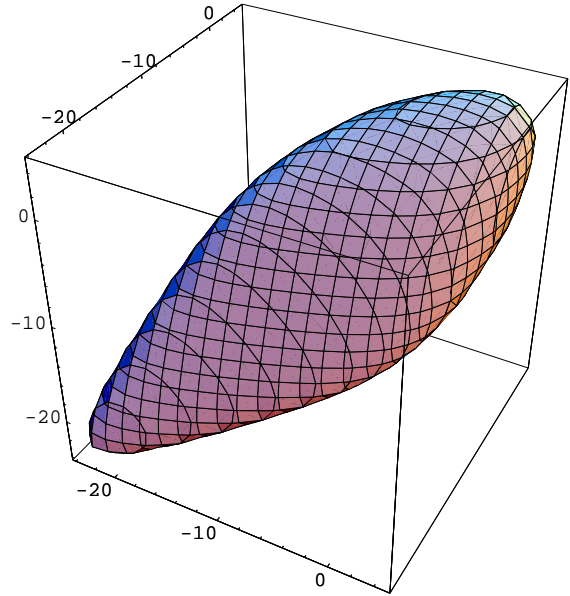}
\end{tabular}
\end{center}
\caption[]{Terms which contribute to contribute  to
$\tilde{\Sigma}(p)$, with $h=0.4$ and $m=p=1$, by more than 1 part in
$10^{10}$ are plotted. To evaluate the sum efficiently, we must focus
on $\{k_1,k_2,k_3\}$ inside this volume.
\label{sunset4}}
\end{figure}

In general, we are faced with the task of computing an $N$-dimensional
infinite sum.  For the case of diagrams which contribute to the
propagator, the general term has the form of \eq{format}, and the
overall sum can be written
\bea
\ILamh(p) &=& \sum_{k} f(k_1,\cdots,k_n) 
 \nonumber \\ 
&=& \sum_{k_1}\left[\cdots\left[\sum_{k_N}
 f(k_1,\cdots,k_n) \right]\cdots\right],
\eea
where $f$ implicitly depends on $m$, $p$, $\Lambda$ and $h$.  We have
focused on the renormalized sunset diagram, but the following
discussion is also applicable to the case with finite $\Lambda$.  When
computing the sums which represent Feynman integrals, it may not be
practical to determine {\em a priori\/} which $\{k_1,\cdots,k_N\}$ are
significant.  Consequently, we have developed a simple algorithm that
evaluates $\ILamh$ by dynamically tracking the size of the general
term as a proportion of the ``running total'' for each of the $N$
nested sums in the overall sum. The precise criteria used to select
the significant terms must be determined heuristically.  The
truncation error must be watched carefully, as all the terms in
$\ILamh$ are positive, and truncating the sum always underestimates
$\ILamh(p)$.  The volume of $\{k_1,k_2,k_3\}$ space which contributes
to $\tilde{\Sigma}(p)$ is shown in \fig{sunset4}.

The $k_i$ only appear in $f(k_1,\cdots,k_N)$ via the $c(k_i)$ and
$p(k_i)$, defined in \eqs{ckdef}{pkdef}.  Because $m^2$ (and
$\Lambda^2$) does not change during the course of a given calculation,
the $c(k)$ and $p(k)$ depend only on the integer values of the
individual $k_i$. Thus we compute $c(k)$ and $p(k)$ for a range of
$k_i$ and storing these results in arrays before starting on the
evaluation of the sum itself. By using these stored values when the
calculating the general term of the sum we avoid the need to recompute
the exponentials in the $p(k)$ and $c(k)$ at every step.

We start each successive sum with an initial value of $k_i$ which
maximized the general term for the previous sum. This is efficient
since $k_{i-1}$ only differs by $\pm1$ between the two cases, and the
dependence of the general term on any of the $k_i$ is sufficiently
weak to ensure that we are starting close to the maximal value.  The
one drawback to this approach is that when computing the final terms
in the sum, we are adding small numbers to a large one (the ``running
total'' for the overall sum), a circumstance that can lead to a loss
of accuracy in fixed precision arithmetic.  We have ensured we are not
losing accuracy in this way by spot-checking our results with
quadruple-precision arithmetic, but in a more sophisticated code this
issue would be addressed more elegantly.

In the case of the sunset diagram, evaluating \eq{sunseth} with $h=0.4$
takes around 3 seconds of workstation CPU time, and the final result
differs from the exact value determined from \eq{sunsetdim} by 1 part
in  $10^{10}$. For $h=0.6$, the accuracy drops to a few parts in $10^8$,
and the execution time drops by around a factor of 3.%
\footnote{More detailed information about the numerical calculations
is given in an appendix. }
For the specific case of the sunset diagram, considerable progress can
be made with the analytic evaluation of the Feynman integral and this
knowledge can be used to produce numerical values for
$\tilde{\Sigma}(p)$.  However, when compared to other methods for {\em
numerically\/} evaluating integrals, the convergence properties and
speed of the Sinc function method is impressive, involving as it does
the direct numerical evaluation of a triple integral.

\subsection{Improvements to Convergence}

While simply adding up all the terms larger than a given minimum size
is the most obvious way to evaluate an infinite sum, it is not
necessarily the most efficient.  In this initial paper we will give
only a brief survey of techniques that could improve the convergence
of the sums.  In general, we foresee three broad classes of technique
for improving convergence; analytical manipulations of the infinite
sum's general term ({\em e.g.\/}
\cite{ArfkenBK1}), numerical extrapolations of successive partial
sums ({\em e.g.\/} \cite{AbramowitzBK1,PressBK1}), and methods that
effectively reduce the order of the $N$ dimensional sums generated by
the Sinc function Feynman rules.

We will not attempt analytical rearrangements or evaluations of the
sums we have derived. However, an interesting line of enquiry would be
to consider whether known analytical treatments of Feynman integrals
have analogous results which can be used to evaluate at least some of
the infinite series in the $\ILamh$.  We do consider one example of
the other two techniques - namely using extrapolation to speed the
evaluation of the innermost sums, and generating a ``dictionary'' of
pre-computed values from which the innermost sum can be accurately
interpolated.

\subsubsection{Aitken Extrapolation}

\begin{figure}[tbp]
\begin{center}
\begin{tabular}{c}
\epsfxsize=8cm 
\epsfbox{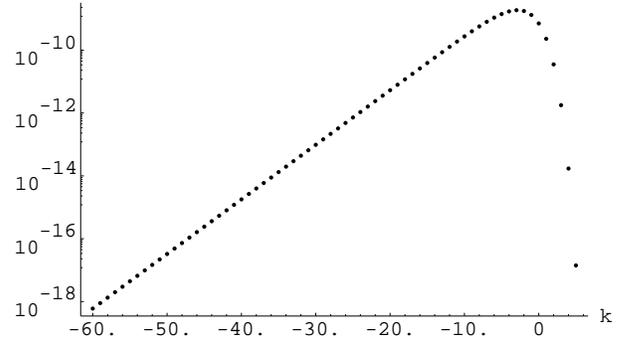}
\end{tabular}
\end{center}
\caption[]{The general term of the Sinc integral form of the
sunset diagram is shown, for $h=0.4$, $m=p=1$ and $k_1=k_2=k_3=k$. The
general term decays like a power-law as $k$ becomes large and
negative.
\label{sunset5}}
\end{figure}

For a general $\ILamh$, if any one of the $k_i$ is becomes large and
positive, the general term is dominated by factor like
$\exp(-e^{k_{i}h})$.  This expression decreases so precipitously that
the transition between terms which are significant and those which are
negligible is very sharp, and there is little to be gained from
extrapolative techniques. However, for $k_i \ll 0$, the general term
takes on a power-law behavior, whose regularity can be exploited by
extrapolation. The general term of the sum is shown (for
$k_1=k_2=k_3$) in \fig{sunset5}.

As a specific example of numerical extrapolation methods, consider the
Aitken $\delta^2$ approach \cite{PressBK1}, which is applicable to
infinite series whose general term behaves like $\sim x^k$ for large
$k$, which is true of our problem when $k_i$ is decreasing. For a
general infinite series with three successive partial sums
$S_{n-1},S_n,S_{n+1}$, Aitken's $\delta^2$ method gives the improved
estimate,
\be \label{aitken1}
S' = S_{n+1} - \frac{(S_{n+1} - S_n)^2}{S_{n+1} - 2 S_n + S_{n-1}}.
\ee
If the general term of the series is $f(n)$, we can rewrite this as
\be \label{aitken2}
S' = S_{n+1} + \frac{f(n+1)^2}{f(n) - f(n+1)}.
\ee
Written in this form, we can recognize the general result for a sum of
the type $\sum_{k=n}^{\infty}x^n$, with $f(n) = x^n$.  For the
specific case of the sunset diagram, using the Aitken $\delta^2$
extrapolation as $k_3$ becomes large and negative can improve the
performance of the algorithm by at least $25\%$.

\subsubsection{Reduction of Order}

As we noted in the previous section, the Sinc function Feynman rules
yield an $N$-dimensional infinite sum, where $N$ is the number of
internal lines in the diagram. Consequently, if we can reduce the
dimension of the infinite sum, we can expect a significant improvement
in efficiency. By considering the general form of the sums derived
from the Sinc function Feynman rules, we can see that while the
overall sum is $N$-dimensional, the summand can typically be specified
by fewer than $N$ independent parameters.  In general, the sums take
the form
\bea
\ILamh &=& \sum_{k_1}\cdots\sum_{k_j}\sum_{k_{j+1}}\cdots \sum_{k_N}
p(k_1) 
\cdots p(k_N)  \times \nonumber \\   &&
\qquad f(d_1,\cdots,d_p,d_{p+1},\cdots,d_q).
\eea
Rearranging the $k_i$ so that $\{k_{j+1},\cdots,k_N\}$ only appear in
a subset, $\{d_{p_1},\cdots,d_{q}\}$ of $\{d_1,\cdots,d_q\}$, we can
write the innermost $N-j$ sums as a function of $\{d_1,\cdots,d_p\}$,
$F(d_1,\cdots,d_p)$, where
\bea
F(d_1,\cdots,d_p) &=& \sum_{k_j+1} \cdots\sum_{k_N} p(k_{j+1})\cdots
 p(k_N) \times \nonumber \\   
&& \qquad f(d_1,\cdots,d_p,d_{p+1},\cdots,d_q).
\eea
Consequently, our sum is now
\be
\ILamh = \sum_{k_1}\cdots\sum_{k_j} p(k_{1})\cdots
 p(k_j) F(d_1,\cdots,d_p).
\ee
This may not seem like an improvement, since we still have to evaluate
the $F(d_1,\cdots,d_p)$. However, by evaluating $F$ for a range of
values of $\{d_1,\cdots,d_p\}$ we can construct an interpolation table
which will yield $F$ to some specified accuracy. Constructing the
interpolation table will typically require $\exp{(c' (N-j+q-p)} $
operations, while the evaluation of the remaining outermost sums will
take a further $\exp{(c''j)}$ operations.  Provided $N-j+p-q$ is less
than $N$, the combined process will most likely be more efficient than
evaluating the $N$ dimensional sum directly.

We now illustrate this process by the applying it to the increasingly
familiar sunset diagram. If we focus on the renormalized quantity
$\tilde{\Sigma}$, we can write \eq{sunseth} as 
\be
\tilde{\Sigma}_h (p) = \frac{m^2 h^3}{(4 \pi)^4} \sum_{k_1,k_2}
p(k_1) p(k_2)   F(d)
\ee
where
\bea
 F(d) &=& \sum_{k_3} \frac{p(k_3)}{\left( d + 1/c(k_3)\right)^2}\
 \left[\exp{ \left(\frac{-p^2}{m^2} \frac{1}{ d + 1/c(k_3)}\right)}
 \right. \nonumber \\ &&\qquad \left. -1 + \frac{p^2}{m^2} \frac{1}{ d
 + 1/c(k_3)}\right],
\\
d &=& \frac{1}{c(k_1)}  + \frac{1}{c(k_2)}.
\eea
In practice, by computing $F(d)$ for several hundred values of $d$ we
calculate $\tilde{\Sigma}(p)$ to within several parts in $10^{10}$
(with $h=0.4$) in less than 1/5th of the time needed for the direct
evaluation of the three dimensional sum.

Another way to speed the evaluation of the diagrams would be to
exploit the symmetry properties of the sums themselves, which reflects
the structure of the underlying diagram. For instance, for a given
$\{k_1,k_2,k_3\}$ the value of the general term in $\tilde{\Sigma}_h$
is not changed by permuting the $k_i$. By using this knowledge it
would be possible to reduce the number of distinct sets of
${k_1,k_2,k_3}$ for which the general term in the sum had to be
computed.

\begin{figure}
\begin{center}
\begin{fmffile}{loop3}
\begin{fmfgraph*}(150,75)
  \fmfpen{thick}        
  \fmfleft{i1} %
  \fmf{fermion,label=$p$}{i1,v1}
  \fmf{fermion,label=$p$}{v3,o1}
  \fmf{vanilla,left,tension=.4}{v1,v2,v1}
  \fmf{vanilla,left,tension=.4}{v3,v2,v3}       
  \fmf{vanilla,left,tension=.2}{v1,v3}          
  \fmfright{o1} 
  \fmfdotn{v}{3}
\end{fmfgraph*}
\end{fmffile}
\end{center}
\caption[]{The three loop diagram that contributes to the $\lambda \phi^4$
propagator, with momentum, $p$. \label{loops3}}
\end{figure}
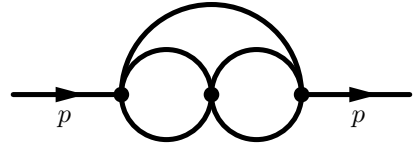

\begin{figure}[tbp]
\begin{center}
\begin{tabular}{c}
\epsfxsize=8cm 
\epsfbox{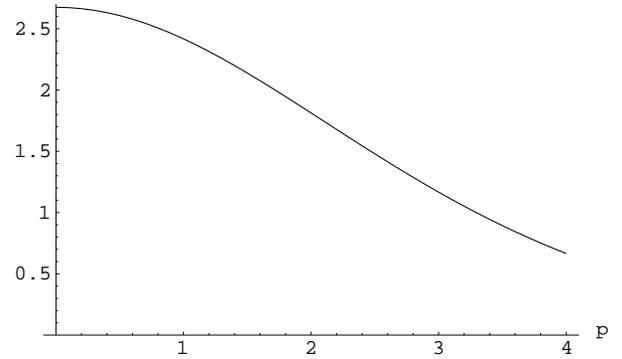}
\end{tabular}
\end{center}
\caption[]{We plot $\SigmaLamh^{(3)}(p)$ for $h=0.8$, $m=1$ and
$\Lambda^2 = 16$. The prefactor of $m^2/(4\pi)^6$ in \eq{sigma3} has
been removed from these numerical results.
\label{vert3}}
\end{figure}

\subsection{Higher Order Diagrams}

Until now, we have used the sunset diagram to illustrate the Sinc
function approach to Feynman integrals.  However, our aim is to
develop a method which applies to arbitrary topologies. We now
evaluate several specific higher order diagrams, although for
convenience, we focus on contributions to the two-point function.

These diagrams typically contain sub-divergences, so we cannot
regularize them using the prescription we employed with the sunset
diagram.  In order to focus on the Sinc function method itself, in
this initial paper we use an explicit cut-off to render the higher
order diagrams finite, and compare the Sinc function results to Monte
Carlo integrations performed with the VEGAS algorithm
\cite{Lepage1978a,PressBK1}.

\subsubsection{Case Study II: Third Order}

Consider the following three loop graph,  

\be
\Sigma^{(3)}(x_1-x_2) = \int d^4 yG(x_1-y)^2 G(x_2-y)^2 G(x_1-x_2),
\ee
whose topology is illustrated in \fig{loops3}. Applying the Sinc
function Feynman rules gives:
\bea
\SigmaLamh^{(3)}(p) &=& \frac{m^2 h^5}{(4\pi)^6} \sum_{k} 
\frac{p(k_1)\cdots p(k_5)}{\left(d_1 d_2 + (d_1+d_2)/c(k_1)\right)^2}
\times
\nonumber\\
&& \exp{\left(-\frac{p^2}{m^2} \frac{d_1 + d_2}{
d_1 d_2 + (d_1 + d_2)/c(k_1)} \right)}, \label{sigma3}
\eea
where
\be d_1 = \frac{1}{c(k_2)} +  \frac{1}{c(k_3)},\qquad
d_2 = \frac{1}{c(k_4)} +  \frac{1}{c(k_5)}.
\ee
As with the sunset diagram, $h$ can be taken close to unity without
inducing a large difference between $\SigmaLam^{(3)}$ and
$\SigmaLamh^{(3)}$, and $\SigmaLam^{(3)}(p)$ can be evaluated to six
or seven significant figures in approximately twenty seconds. Since we
have no independent estimate of $\SigmaLam^{(3)}(p)$, we heuristically
determine the number of significant figures in our result by varying
both $h$ and the size of the largest discarded terms in the multiple
sum.  Hopefully, a more careful analysis will give an {\em a priori\/}
understanding of the truncation error in the evaluation of the series,
and the $h$ dependence of the error in $\ILamh$.

The results obtained from \eq{sigma3} can be independently checked by
using VEGAS to evaluate the five dimensional integral representation
of $\SigmaLam^{(3)}$. As well as confirming the Sinc function
calculation, this also allows us to compare the time required to
evaluate the Sinc function representation with the Monte Carlo
algorithm.  We find that to obtain four or five significant figures of
$\Sigma^{(3)}$ using VEGAS takes around 50 times longer than it took
us to find seven significant figures of $\Sigma^{(3)}$ from $\ILamh$.

We do not place wish to place too much weight on this comparison.  In
practice, it is very unlikely that any $\ILam$ would be evaluated
directly, as it is almost always possible to perform some analytical
simplifications.  On the other side of the ledger, we have also not
attempted make analytical improvements to $\ILamh$.  Moreover, VEGAS
is a mature and well-tested algorithm. Conversely, we are using a very
simple approach to evaluate the $\ILamh$, and the previous results
suggest that a more sophisticated algorithm would significantly reduce
the time needed to evaluate the sums.  Finally, the $\ILam$ are, in
general, cut-off dependent, whereas one usually wishes to calculate
cut-off independent quantities. These can be combinations of different
$\ILam$ and their derivatives, in the limit that $\Lambda\rightarrow
\infty$.  This process of extracting cut-off independent quantities is
presumably similar for the Sinc function and integral forms, but will
generally lead to both integrands and summands which are more
complicated that those which appear in the $\ILam$ and $\ILamh$.

However, while they are based on an artificial comparison, these
results do suggest that Sinc function methods may lead to an useful
approach for evaluating Feynman diagrams, which is intrinsically
faster than one which relies on the use of Monte Carlo integration.
Moreover, any advantage will become more pronounced when significant
levels of numerical precision are required.

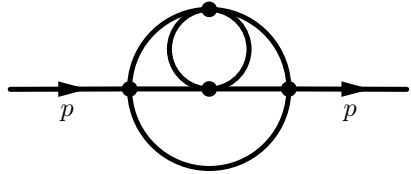
\begin{figure}
\begin{center}
\begin{fmffile}{loop4}
\begin{fmfgraph*}(150,80)
  \fmfpen{thick}        
  \fmfleft{i1} %
  \fmf{fermion,label=$p$}{i1,v1}
  \fmf{fermion,label=$p$}{v2,o1}
  \fmf{vanilla}{v1,v3,v2}
  \fmfforce{75,70}{v4}
  \fmfforce{45,40}{v1}
  \fmfforce{105,40}{v2}   
   \fmf{vanilla,right,tension=0}{v1,v2,v1}
  \fmfright{o1} 
\fmffreeze
  \fmf{vanilla,left,tension=0}{v3,v4,v3}
  \fmfdotn{v}{4}
\end{fmfgraph*}
\end{fmffile}
\end{center}
\caption[]{A four loop diagram that contributes to the $\lambda \phi^4$
propagator, with momentum, $p$. \label{loops4}}
\end{figure}

\begin{figure}[tbp]
\begin{center}
\begin{tabular}{c}
\epsfxsize=8cm 
\epsfbox{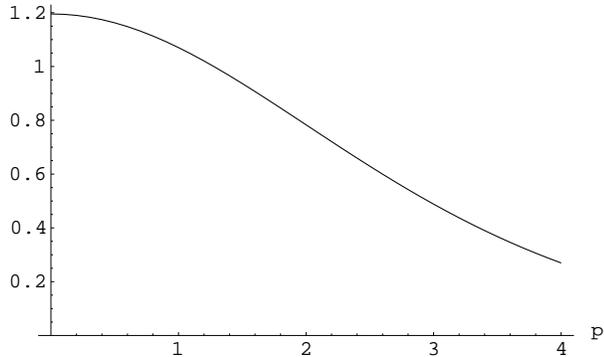}
\end{tabular}
\end{center}
\caption[]{We plot $\SigmaLamh^{(4)}(p)$ for $h=0.9$, $m=1$ and
$\Lambda^2 = 16$. As is the case with the 3-loop example we consider,
this quantity is explicitly cut-off dependent, and we have
suppressed the $m^2/(4\pi)^8$ prefactor that appears in the Sinc
function representation.
\label{vert4}}
\end{figure}

\subsubsection{Case Study III: Fourth Order}

The final diagram we consider here is the four loop contribution to the
propagator, shown in \fig{loops4}, which can be represented as the
following integral over propagators,
\bea
\Sigma^{(4)}(x_1-x_2) &=& \int d^4 y_1 d^4 y_2 
   G(x_1-x_2) G(y_1-y_2)^2 \times \nonumber\\
&& \qquad G(x_1-y_1) G(x_1-y_2)\times \nonumber\\
&& \qquad G(x_2-y_1) G(x_2-y_2).
\eea
It is again straightforward to apply the Sinc function Feynman rules
and take the Fourier transform to find $\SigmaLamh^{(4)}(p)$, and the
results of evaluating it for a specific set of $\Lambda$, $m$ and $p$
are shown in \fig{vert4}.

In this case, VEGAS takes around 20 minutes to find a result to within
1 part in $10^4$, whereas the Sinc function method returns a value
accurate to 1 part in $10^6$ in approximately five minutes. Again, the
Sinc function approach outperforms VEGAS by around two orders of
magnitude, although the seven-dimensional sum is appreciably more
expensive to evaluate than the previous five-dimensional example.

\section{Conclusions and Discussion}

In this paper we have developed the Sinc function representation,
which associates arbitrary Feynman integrals with multi-dimensional,
infinite sums.  The basis of this representation is an approximation
to the propagator, derived using the theory of generalized Sinc
functions. We have used the Sinc function representation to develop a
new approach to numerically evaluating Feynman integrals. This method
is simple to implement and is potentially faster than Monte Carlo
algorithms, which are currently used to evaluate most Feynman
integrals that cannot be performed analytically.

We presented three ``case studies'' which illustrate the properties of
the Sinc function representation.  Firstly, we evaluated the two loop
sunset diagram, and compared the Sinc function representation to
explicit evaluations of the corresponding integral. We showed that we
can easily compute results that are accurate to within a few parts in
$10^{10}$, and that for diagrams with no sub-divergences it is
straightforward to extract regularization independent quantities from
the Sinc function representation.  We then evaluated representative
third and fourth order diagrams from the propagator expansion of
$\lambda\phi^4$ field theory.

For the three and four loop diagrams we evaluated, we contrasted the
Sinc function representation with direct calculations of the
corresponding integrals with VEGAS, showing that the Sinc function
representation is significantly more efficient than VEGAS. This
advantage increases as the desired level of accuracy is raised, and
can easily amount to several orders of magnitude in execution time.
However, this direct comparison between Monte Carlo and Sinc function
methods is somewhat artificial since the quantities computed are
cut-off dependent, and we did not attempt to analytically simplify the
integrals (or the sums) before evaluating them.  Consequently, it
remains to be seen whether the Sinc function representation will speed
up realistic calculations.

In order to test the Sinc function representation in practical
situations, we plan to investigate a number of topics.  We have
already used the Sinc function representation to evaluate the
three-loop master diagrams whose analytical properties are discussed
in detail by Broadhurst \cite{Broadhurst1998a}. In the course of this
work \cite{EastherET1999e} we explicitly extended the Sinc function
representation to diagrams with massless lines, and reproduced the
analytical values of the master diagrams to better than 1 part in
$10^{10}$.  In addition to describing a renormalization procedure for
the Sinc function representation of arbitrary diagrams, our immediate
priority is to include fermion and vector fields, and thus to evaluate
diagrams from QED and electroweak theory.  We have calculated several
simple QED diagrams with Sinc function techniques \cite{Wang1998a},
and we are currently working on the systematic application of the Sinc
function representation to arbitrary QED diagrams.

We have used a straightforward procedure to evaluate the
multi-dimensional sums derived from the Sinc function Feynman rules.
In Section 4 we discussed a variety of approaches to speeding up the
computation of the sums, and it is possible that the numerical
performance of the Sinc function representation can be significantly
improved.  Moreover, we have not yet attempted to analytically
simplify Sinc function representations before evaluating then
numerically. Consequently, the CPU time we currently need to evaluate
the sum only provides an upper limit on what can be achieved.

Obtaining the Sinc function representation for a specific topology
amounts to evaluating one or more Gaussian integrals, and the process
is easily automated with any computer algebra package.  If the Sinc
function representation is useful in realistic situations, we expect
it will be straightforward to integrate it with existing tools for
automatically evaluating diagrams.  Moreover, applying the Sinc
function Feynman rules to diagrams which contain several fields with
different masses is no more complicated than evaluating the same
topology with identical lines. Diagrams with more than two external
lines also present no new difficulties, although in this case more
than one Fourier transform is needed to move from coordinate to
momentum space, since an $n$-point diagram has $n-1$ independent
external momenta.

Strictly speaking the Sinc function representation is not just a
``better way to do integrals'', since when evaluated to arbitrary
precision, the Sinc function representation converges to an
approximation to the integral, rather than the integral itself.
However, this approximation can be made arbitrarily accurate, so in
practice evaluating the Sinc function representation is equivalent to
a direct computation of the integral. Moreover, several other
approaches lead to expressions for Feynman integrals that are written
as infinite sums
\cite{Mendels1978a,ChetyrkinET1980a,Terrano1980a,BroadhurstET1995a}.
However, these methods yield expressions for the exact integrals, and
are therefore not equivalent to the Sinc function representation.

While many approaches to numerical integration outperform Monte Carlo
techniques in one or two dimensions, their efficiency scales typically
very poorly with the dimension of the integral. For the integrals
derived from Feynman diagrams, the Sinc function representation
provides an exception to this general rule.  For example, the 4th
order diagram we studied can be expressed as a seven dimensional
integral, but a Monte Carlo integration with VEGAS takes much longer
than the Sinc function representation to achieve the same level of
accuracy.  The task before us now is to harness the mathematical
properties of the Sinc function representation, and apply them to
solving physical problems.

\appendix

\section{Numerical Details}  
The numerical results described in this paper we all obtained on the
same Sun workstation, with 250MHz Ultrasparc II CPUs. We refrain from
giving precise timings, since these depend strongly on the specific
combination of hardware and software, and the timings we do give
reflect the use of aggressive compiler optimization settings.  The
codes are implemented in Fortran 77. Sample codes (including those
used to perform the calculations described in this paper) are
available at the following URL:\\
{\tt http://www.het.brown.edu/people/easther/feynman}

\section*{Acknowledgments}

We thank Pinar Emirda\u{g}, Toichiro Kinoshita, and Wei-Mun Wang for
useful discussions.  Computational work in support of this research
was performed at the Theoretical Physics Computing Facility at Brown
University.  This work is supported by DOE contract DE-FG0291ER40688,
Tasks A and D.


\begin{thebibliography}{10}

\bibitem{Lepage1978a}
G.~P. Lepage, J. Comput. Phys. {\bf 27},  192  (1978).

\bibitem{PressBK1}
W. Press, S. Teukolsky, W. Vetterling, and B. Flannery, {\em Numerical Recipes
  in Fortran}, 2 ed. (Cambridge UP, Cambridge, 1992).

\bibitem{Kinoshita1990a}
T. Kinoshita,  in {\em Quantum Electrodynamics}, edited by T. Kinoshita (World
  Scientific, Singapore, 1980), pp.\ 218--321.

\bibitem{HarlanderET1998a}
R. Harlander and M. Steinhauser, hep-ph/9812357  (1998).

\bibitem{GarciaET1994a}
S. Garc\'{\i}a, G.~S. Guralnik, and J.~W. Lawson, Phys. Lett. B {\bf 333},  119
   (1994).

\bibitem{GarciaET1996a}
S. Garc\'{\i}a, Z. Guralnik, and G.~S. Guralnik, hep-th/9612079  (1996).

\bibitem{LawsonET1996a}
J.~W. Lawson and G.~S. Guralnik, Nucl. Phys. B {\bf 459},  589  (1996).

\bibitem{Hahn1998a}
S.~C. Hahn, Ph.D. thesis, Brown University, 1998.

\bibitem{HahnET1999a}
S.~C. Hahn and G.~S. Guralnik, hep-th/9901019  (1999).

\bibitem{StengerBK1}
F. Stenger, {\em Numerical Methods Based on Sinc and Analytic Functions}
  (Springer Verlag, New York, NY, 1993).

\bibitem{Higgins1985a}
J.~R. Higgins, Bull A. M. S. {\bf 12},  45  (1985).

\bibitem{RamondBK1}
P. Ramond, {\em Field Theory: A Modern Primer}, 2 ed. (Addison Wesley, Redwood
  City, CA, 1989).

\bibitem{ArfkenBK1}
G. Arfken, {\em Mathematical Methods for Physicists}, 3 ed. (Academic Press,
  Orlando, 1985).

\bibitem{AbramowitzBK1}
{\em Handbook of Mathematical Functions}, edited by M. Abramowitz and I. Stegun
  (Dover, New York, NY, 1965).

\bibitem{Broadhurst1998a}
D. Broadhurst, Eur. Phys. J. {\bf C8},  311  (1999).

\bibitem{EastherET1999e}
R. Easther, G. Guralnik, and S. Hahn, \\hep-ph/9912255  (1999).

\bibitem{Wang1998a}
W.-M. Wang, Perturbative calculations in quantum field theory, ScB Thesis,
  Brown University, 1998.

\bibitem{Mendels1978a}
E. Mendels, Nuovo Cimento {\bf 45A},  87  (1978).

\bibitem{ChetyrkinET1980a}
K.~G. Chetyrkin, A.~L. Kataev, and F.~V. Tkachov, Nuc. Phys. B {\bf 174},  345
  (1980).

\bibitem{Terrano1980a}
A.~E. Terrano, Phys. Lett. B {\bf 93},  424  (1980).

\bibitem{BroadhurstET1995a}
D.~J. Broadhurst and D. Kreimer, Int. J Mod. Phys. C {\bf 6},  519  (1995).

\end{thebibliography}

\end{document}